\documentclass{frontiersFPHY}
\usepackage{url,lineno}
\usepackage{amssymb}
\usepackage{amsfonts,color}

\copyrightyear{}
\pubyear{}

\def\journal{Physics}%%% write here for which journal %%%
\def\DOI{}
\def\articleType{Research Article}
\def\keyFont{\fontsize{8}{11}\helveticabold }
\def\firstAuthorLast{Cal\c{c}ada {et~al.}} %use et al only if is more than 1 author
\def\Authors{Marcos Cal\c{c}ada\,$^{1}$, Jos\'e T. Lunardi\,$^{1,*}$, Luiz A. Manzoni\,$^{2}$ and Wagner Monteiro\,$^1$}
\def\Address{$^{1}$Departamento de Matem\'atica e Estat\'{\i}stica,
Universidade Estadual de Ponta Grossa, Ponta Grossa, PR, Brazil \\
$^{2}$Department of Physics, Concordia
College, Moorhead, MN, USA }
\def\corrAuthor{Jos\'e T. Lunardi}
\def\corrAddress{Departamento de Matem\'atica e Estat\'{\i}stica, Universidade Estadual de Ponta Grossa. Avenida General Carlos Cavalcanti, 4748. 84030-900, Ponta Grossa, PR, Brazil}
\def\corrEmail{jttlunardi@uepg.br}

\begin{document}
\onecolumn
\firstpage{1}

\title[Distributional approach to point interactions]{Distributional approach to point interactions in one-dimensional quantum mechanics}
\author[\firstAuthorLast ]{\Authors}
\address{}
\correspondance{}
\extraAuth{}% If there are more than 1 corresponding author, comment this line and uncomment the next one.
%\extraAuth{corresponding Author2 \\ Laboratory X2, Institute X2, Department X2, Organization X2, Street X2, City X2 , State XX2 (only USA, Canada and Australia), Zip Code2, X2 Country X2, email2@uni2.edu}
\topic{}% If your article is part of a Research Topic, please indicate here which.

\maketitle

\begin{abstract}
We consider the one-dimensional quantum mechanical problem of defining interactions concentrated at a single point in the framework of the theory of distributions. The often ill-defined product which describes the interaction term in the Schr\"odinger and Dirac equations is replaced by a well-defined distribution satisfying some simple mathematical conditions and, in addition, the physical requirement of probability current conservation is imposed. A four-parameter family of interactions thus emerges as the most general point interaction both in the non-relativistic and in the relativistic theories (in agreement with results obtained by self-adjoint extensions). Since the interaction is given explicitly, the distributional method allows one to carry out symmetry investigations in a simple way, and it proves to be useful to clarify some ambiguities related to the so-called $\delta^\prime$ interaction.

\tiny
 \keyFont{ \section{Keywords:} Quantum mechanics in one dimension, point interactions, distribution theory, $\delta$-interaction, $\delta '$-interaction} %All article types: you may provide up to 8 keywords; at least 5 are mandatory
\end{abstract}

\section{Introduction}
\label{intro}

Point (zero-range) interactions have attracted great interest in quantum mechanics \cite{AGH04, CNT99, SCL02, GNN09, ACG11, CNP97, Zol10}. They provide important solvable models with a wide variety of applications in atomic
physics, such as the Lieb-Liniger \cite{LLi63, Lie63} model for a one dimensional gas of bosons interacting by means of a
$\delta$-function potential (for other applications see, e.g., \cite{Exn95,Kun99,CTT10} and references therein). In addition, point
interactions have proven to be a fruitful \emph{theoretical laboratory} to investigate methods of quantum field
theory (QFT), such as regularization and renormalization, in simpler and more manageable models in one and
two-dimensional quantum mechanics \cite{Jac91, MTa94, Nye00}.

The simplest point interaction in one dimensional non-relativistic quantum mechanics is given by the Dirac's $\delta$-function potential, which is well defined and has well-known solutions. However, attempts to consider more general interactions, such as that associated with a $\delta '$ potential (the prime
indicates a spatial derivative), have been known to be plagued with difficulties associated with the definition of the interaction \cite{CNP97,Seb86, Zha92,Gri93,AGH93, RTa96, CSh98} (also see \cite{GWa99} and references therein).

Mathematically rigorous methods were employed to define such interactions by obtaining self-adjoint extensions
(SAE) of the kinetic energy operator for Schr\"{o}dinger's theory. It was shown that a four-parameter family of interactions, defined by their boundary conditions, exhaust all the possibilities for point interactions in non-relativistic quantum mechanics \cite{CNP97,Seb86b, Exn96} (for a review of the mathematical literature on this subject see \cite{AGH04}).

Given the somewhat abstract and mathematically involved nature of SAE, another approach was developed for this subject, consisting in the investigation of regularizations using sequences of regular short-range potentials which converge to point potentials in the zero-range limit \cite{Seb86, RTa96, CSh98, Car93}. Even if this method is more intuitive and appealing from a physicist's point of view, such an approach has often led to ambiguous and even contradictory results -- which, in general, arise from the dependence of the particular regularizing scheme employed (see, for instance, \cite{ACG11, Zol10,CAZ03} and references there cited).

In relativistic quantum mechanics, for a Dirac particle, even the $\delta$-function potential (which is the most singular interaction allowed in the Dirac's equation) is problematic and it is
known to lead to ambiguities \cite{SMa81, MSt87, DAM89, Roy93}. Similarly to the non-relativistic case, the SAE approach implies that the most general point interactions in Dirac's theory are also
given by a four-parameter family of interactions \cite{FGR87, BDa94}. The most commonly
studied particular cases of relativistic point interactions correspond to pure electrostatic and pure scalar
potentials, which have also been investigated by regularization through $\delta$-converging sequences of
functions, with results that are, in general, dependent on the regularizing function used \cite{ACG11, SMa81,
MSt87}.

The situation described above is not entirely new: analogous difficulties involving regularization ambiguities
arise in quantum field theory (QFT) and have been successfully addressed by using the Epstein-Glaser approach to
QFT, in which distribution theory is employed from the beginning and the symmetries of the system are carefully
implemented (see, e.g., \cite{EGl73,Sch95, MPT99,LPM02} and references therein). In a similar vein, in this paper we will introduce a distributional approach to point interactions in quantum mechanics -- first announced in a preliminary version in \cite{LMM13}.

The method proposed here starts from the realization that both the singular potential and the wave function in the Schr\"odinger or Dirac equations \emph{must} be considered as distributions. This precludes the naive definition of the interaction term in these equations as the usual product between the wave function and the potential energy, since such a product is generally ill-defined in distribution theory \cite{Zem87, GSh}. We will introduce a well defined distribution to play the role of the interaction term, the properties of which must follow from simple mathematical requirements concerning its support and singular order (a concept introduced in the next section), besides the physical requirement of conservation of the probability current across the singular point. As a consequence, a four-parameter family of distributions will naturally emerge as the most general point interaction in both non-relativistic and relativistic one-dimensional quantum mechanics, a result which is in complete agreement with those obtained with the SAE approach \cite{BDa94, ADK98}. However, there is an important conceptual difference between the SAE and the distributional approaches, for while the first characterizes all the possible point interactions only through the boundary conditions that the wave function must satisfy around the singular point, the latter explicitly gives the interaction term as a distribution concentrated at that point -- thus, in the distributional approach the boundary conditions emerge as a \emph{consequence} of the properties of the particular interaction distribution substituted into the Schr\"odinger or Dirac equations. Thus, the distributional approach provides an alternative (also mathematically rigorous) to the method of SAE, with the advantage of being less abstract (to a physicist) and allowing a more intuitive physical interpretation of the interaction term, which here is presented explicitly as a distribution. In addition, the distributional approach can contribute to clarify some ambiguities which arise from the use of regularizations to treat zero-range potentials -- in particular, the explicit form of the interaction distribution allow us to address, in a straightforward way, the much debated question about the existence or not of a ``true" (i.e., \emph{odd} under parity) point interaction associated with a $\delta'$ potential. Notice that a proper definition of the notion of an odd point interaction is a non-trivial task in the context of the SAE approach.

It is worth to notice that in this work we are concerned only to the well-established Schwartz's
theory of distributions, which considers only distributions defined on spaces of infinitely differentiable test functions. An approach based on an alternative theory of distributions defined
on spaces of discontinuous test functions was also proposed by Kurasov in \cite{Kur96}, who assumes the wave function to be a member of a space of test functions and claims that point interactions cannot be properly defined in the framework of the standard distribution theory. In the approach proposed here such a difficulty does not arise because the wave function itself must be consistently treated as a distribution, not as a test function. This is analogous to what occurs in the Epstein-Glaser distributional approach to quantum field theory, in which quantities defined in terms of quantum fields must be treated themselves as (operator-valued) distributions \cite{EGl73,Sch95}.

This paper is organized as follows. In section \ref{NR} we introduce the distributional approach by considering the Schr\"{o}dinger equation with a singular potential of order $r_s=1$, the most singular point interaction allowed in non-relativistic one-dimensional quantum mechanics. In Section \ref{R} we extend our methods to consider Dirac's equation with a singular interaction of order $r_s = 0$. Section \ref{concl} is reserved for our concluding remarks.
%%%%%%%%%%%%%%%%%%%%%%%%%%%%%%%%%%%%%%%%%%%%%

\section{Schr\"{o}dinger's equation with point interactions}
\label{NR}

In this Section we introduce the distributional approach to point interactions in one dimensional quantum mechanics. We consider distributions defined on the Schwartz space of rapid descent test functions -- for a comprehensive treatment of distributions see \cite{Zem87}, which we follow closely (also see \cite{GSh}).

For a non-relativistic particle moving under the influence of a  potential energy $V(x)$ which is assumed to be a locally integrable function (i.e., a function integrable in the Lebesgue sense over every finite interval), the one-dimensional time-independent Schr\"odinger equation can be written as (in this section we will adopt units such that $\hbar =1$ and $2m=1$)
\begin{equation}
\label{sch}
\frac{d^2}{dx^2} \psi(x) + E \psi (x) = V(x)\psi(x) \, .
\end{equation}
Then, by assuming that the wave function $\psi$ is continuous everywhere, the \emph{interaction term}, $V(x)\psi(x)$, is well-defined and corresponds to a locally integrable ordinary function. On the other hand, potentials which do not correspond to locally integrable functions are called \emph{singular}, as it is the case of the Dirac $\delta$-``function" or its derivatives, and only make sense as distributions. Distributions corresponding  to locally integrable functions are said to be \emph{regular}. As a consequence, in the case of singular potentials the Schr\"odinger equation is in general mathematically ill-defined, since the wave function itself must be considered as a distribution and the product appearing on the r. h. s. of  (\ref{sch}) is not a well defined operation for two arbitrary distributions \cite{Zem87,GSh}.

The starting point of a distributional approach to singular interactions is to rewrite Schr\"{o}dinger's equation as
\begin{equation}
\psi ''(x)+k^2\psi (x)=s[\psi](x) \, ,
\label{Ss}
\end{equation}
where $k \equiv \sqrt{E}$, the prime indicates a distributional derivative with respect to $x$, and the interaction term was substituted by a distribution $s[\psi](x)$, still to be determined by mathematical and physical requirements, and which univocally determines the interaction. Then, in any interval which does not include the singularities the potential is regular (and the wave function $\psi$ continuous), and therefore the interaction distribution, $s[\psi](x)$, coincides with the ordinary product of functions $\psi(x)V(x)$ in such an interval.

In order to determine the interaction distribution $s[\psi](x)$ let us define the \emph{singular order} (or simply the \emph{order}) of a distribution, a concept that characterizes the ``strength" of the singularity and plays a crucial role in this approach. The definition of singular order of a distribution adopted in this work is the extended concept of order considered by Zemanian (see \cite{Zem87}, p. 162-163), which can be written as follows. A distribution $f$ is said to have singular order $r\in \mathbb{Z}$ on a closed finite interval $I$ if $f=v^{(r+2)}$ in this interval, where $v$ is a distribution corresponding to a continuous but not differentiable function on $I$ and $v^{(r+2)}$ is the $(r+2)$-th distributional derivative of $v$ (a ``derivative of negative order" means an indefinite integration, as usual). An infinitely smooth function is said to have order $r=-\infty$. If a distribution has a finite order the operations of differentiation and indefinite integration change its order by $+1$ and $-1$, respectively. It follows that distributions having $r\leq -2$ are regular, whereas distributions having $r\geq 0$ are singular. When $r=-1$ the distribution can be regular or singular, depending on the case. As some important examples, the Heaviside $\theta$ distribution has singular order $-1$ (and it is a regular distribution), while its primitives have $r=-2$, and correspond to continuous but not differentiable functions. The Dirac's delta $\delta (=\theta')$ and its first derivative $\delta'$ are singular distributions with singular orders $0$ and $+1$, respectively.

Let us now return to the task of determining the interaction distribution $s[\psi](x)$. We require that:
\begin{itemize}

\item[($i$)] $\mathrm{supp}\; s\subseteq \mathrm{supp} \; V$;

\item[($ii$)] the singular order of the distribution $s[\psi ](x)$ cannot exceed the singular order of $V(x)$ in any closed subset of its support;

\end{itemize}
where $\mathrm{supp}\,s$ indicates the \emph{support} of the distribution $s$, which is the smallest closed set outside of which $s$ equals zero \cite{Zem87}. Both these requirements are automatically satisfied when $V(x)$ corresponds to a regular distribution -- here we are just extending them in a natural way to the case of potentials described by singular distributions. In particular, requirement $(ii)$ prevents the introduction of interaction terms even more singular than the original potential (this is analogous to the \emph{minimal distribution splitting} condition introduced in the Epstein-Glaser approach to QFT \cite{EGl73,Sch95}).

In this paper we are interested only in singular \emph{point interactions}, i.e., interactions which vanish everywhere except at a single point, which is assumed to be the origin $x=0$, where they are singular. Thus, from the requirement $(i)$ above, it follows that $\mathrm{supp}\; s[\psi]=\{0\}$.

We now recall a fundamental theorem from distribution theory (see \cite{Zem87} Theorem 3.5-2, p. 98):

\begin{quotation}
\noindent
\textbf{Theorem 1.} A necessary and sufficient condition for a distribution $f(x)$ on $\mathbb{R}$ to have a support
concentrated at the origin is that it be a finite sum %
\begin{equation}
\label{cldelta}
   f(x)\,=\,\sum_{m=0}^{r_s} \,\alpha_m \,\delta^{(m)}(x)\, ,
\end{equation}
where the $\alpha_m$ are (complex) constants, $\alpha_{r_s}\neq 0$, $\delta^{(m)}$ is the $m$-th derivative of the $\delta$ distribution, and $r_s$ is the singular order of
distribution $f$.
\end{quotation}

From this theorem we conclude that the interaction distribution $s[\psi](x)$ must be a linear combination of the delta distribution and its derivatives at the origin, which implies
\begin{equation}
\label{clinterm}
   s[\psi](x)\,=\,\sum_{m=0}^{r_s} \,\alpha_m[\psi] \,\delta^{(m)}(x)\, ,
\end{equation}
where $r_s$ stands for the \emph{singular order} of the interaction term,  $\delta^{(m)}$ denotes the $m$-th derivative of the $\delta$-distribution and $\alpha_m[\psi]$ are complex numbers, expressed as \emph{linear functionals} depending  on the behavior of the wave function \emph{around the singular point}, i.e., $\alpha_m[\psi]\equiv \alpha_m[\psi\left(0^\pm\right),\psi'\left(0^\pm\right)]$. This last condition is needed if only \emph{local} interactions are to be considered, and the linearity of $\alpha_m[\psi]$ is required by the superposition principle.

The specification of the singular order $r_s$ and of all the coefficients $\alpha_m$ ($m=0,\cdots,r_s$) in (\ref{clinterm}) uniquely determines the interaction. However, in order to determine such quantities we must impose on the system some fundamental physical requirements and, since we are considering a \emph{stationary system}, it is natural to require that
\begin{itemize}

\item[$(iii)$] the probability current must be conserved everywhere -- in particular, it must be conserved across the singular point;

\end{itemize}
There is, of course, nothing new about this requirement in itself -- however, when considered together with $(i)$ and $(ii)$ it drastically reduces the number of free parameters in a point interaction, as shown below.

From now on, let us restrict ourselves to point interactions of singular order $r_s = 1$ and look for the most general $s[\psi](x)$ allowed by the requirements $(i)$-$(iii)$. We observe that $r_s = 1$ is the maximum singular order allowed in order to have normalizable wave functions (i.e. square integrable ordinary functions) -- if $r_s>1$ were allowed the stationary wave function would have singular order greater than or equal to zero, and it would necessarily be a singular distribution.

An example of a potential with singular order $r_s = 1$ is given by the so-called delta prime potential energy, namely $V(x) = \gamma\delta'(x)$ with the strength of the interaction $\gamma$ being a real constant. This potential, when naively substituted in Schr\"odinger's equation (\ref{sch}), is well-known to lead to contradictions -- see, e.g., \cite{CNP97, AGH93,GWa99}. The reason for such contradictions is that the product $\psi (x) \delta'(x) $ can be defined in distribution theory if, \emph{and only if}, both the wave function \emph{and} its first derivative \emph{are continuous at the origin} (in which case it is $\psi(x) \delta'(x)=-\psi'(0)\delta(x)+\psi(0)\delta'(x)$, see \cite{Zem87}) -- but such requirement is not compatible with Schr\"{o}dinger's equation (\ref{sch}), as a simple analysis of the singular order shows: both sides of this equation must have the same singular order at the origin, and the r.h.s. $-\psi'(0)\delta(x)+\psi(0)\delta'(x)$ has singular order $+1$; hence $\psi ''(x)$, $\psi '(x)$ and $\psi (x)$ must have singular orders $+1$, $0$, and $-1$, respectively, which means that \emph{the wave function must be discontinuous} at the origin.

Let us return to our task of determining the $s[\psi](x)$ which corresponds to a point interaction with singular order $r_s = +1$. According to conditions ($i$), ($ii$) and Theorem 1, one must consider (\ref{Ss}) for the most general interaction distribution with singular order $+1$ and  concentrated at the origin:
\begin{equation} \label{sp}
s\left[\psi \right](x) = \alpha_0\left[\psi \right] \delta(x) + \alpha_1\left[\psi \right] \delta '(x)\, ,
\end{equation}
with $\alpha_1 \left[\psi \right] \neq 0$. The coefficients $\alpha_0 \left[\psi \right]$ and $\alpha_1\left[\psi \right]$ have yet to be determined (the $\delta$-function potential, which has singularity $r_{\delta} = 0$, can be obtained as a particular case of the above interaction by requiring $\alpha_0\left[\psi \right] \neq 0$ and $\alpha_1 \left[\psi \right] = 0$).

We recall that taking the indefinite integral of a distribution decreases its singular order by one. Thus, since $\psi$ must have singular order $-1$, any primitive $\psi^{(-1)}$ must have singular order $-2$, and corresponds to a function continuous (but not differentiable) at the origin. Therefore, by taking the indefinite integral in both sides of eq. (\ref{Ss}), with the interaction distribution given by (\ref{sp}), we have
\begin{equation}
\label{indint}
\psi'(x)+k^2\psi^{(-1)}(x)=\alpha_0\left[\psi \right] \theta(x)+\alpha_1 \left[\psi \right]\delta(x)+c_1,
\end{equation}
where $c_1$ is an arbitrary (complex) constant. In any interval which does not include the origin both sides of the above equation equal an ordinary function (in particular, the distribution $\delta$ vanishes identically in such an interval). Taking into account the continuity of $\psi^{(-1)}$ at the origin, from the above expression we have
\begin{equation}
\label{bc1}
\psi'\left(0^+\right)-\psi'\left(0^-\right)=\alpha_0 \left[\psi \right].
\end{equation}
Similarly, taking the indefinite integral of (\ref{indint}), and using the continuity of any primitive $\theta^{(-1)}$ of the distribution $\theta$ (since such a primitive has singular order $-2$), we conclude that
\begin{equation}
\label{bc2}
\psi\left(0^+\right)-\psi\left(0^-\right)=\alpha_1 \left[\psi \right].
\end{equation}
Equations (\ref{bc1}) and (\ref{bc2}) reflect the fact that the boundary conditions (b.c.) at the origin are completely determined from the knowledge of the interaction, i.e., from the functional coefficients $\alpha_0$ and $\alpha_1$.

The above b.c. can be written in terms of the wave function and its derivative by noticing that the most general way to express the linear functionals $\alpha_0$ and $\alpha_1$ is
\begin{equation}
 \Omega \equiv \label{funct}
\left( \!\!
\begin{array}{c}
\alpha_1\\
\alpha_0
\end{array} \!\!
\right)=M_+ \Phi\left(0^+\right) - M_- \Phi\left(0^-\right), \qquad {\rm with} \qquad \Phi (x)\equiv
\left( \!\!
\begin{array}{c}
\psi (x)\\
\psi'(x)
\end{array} \!\!
\right)
\end{equation}
where $M_{\pm}$ are $2\times 2$ complex matrices yet to be determined.

Equation (\ref{funct}) is a direct consequence of requirements $(i)$ and $(ii)$ and the solution of the \emph{distributional} Schr\"odinger equation with the general interaction (\ref{sp}). This alone, however, does not impose any constraints on the form of $M_{\pm}$ (i.e., on the functionals $\alpha_0$ and $\alpha_1$); such constraints will come from imposing, in addition to $(i)$ and $(ii)$, condition $(iii)$.

Conservation of the probability current across the origin simply means $j\left(0^-\right)=j\left(0^+\right)$ and, since in any interval which does not include the origin the distribution describing the probability current coincides with the ordinary product of functions $j(x)= -i \left[ \psi^*(x) \psi'(x)- {\psi^*}'(x) \psi(x)\right]$, both sides of that equation are well defined and finite for a general point interaction.

In order to impose $(iii)$ it is convenient first to use the identity $|z+i w|^2-|-z+iw|^2= \frac{2}{i} \left(z w^*-z^*w \right)$, which is valid for any pair of complex numbers $z$ and $w$, and rewrite the current density as \cite{FTs00}
\begin{equation}\label{current}
    L_0 j(x)= \frac{1}{2} \left[ |L_0 \psi'(x) +i \psi(x)|^2-  |- L_0 \psi'(x)+i \psi(x)|^2  \right],
\end{equation}
where $L_0$ is an arbitrary non-zero constant inserted for dimensional reasons. Introducing the vectors
\begin{equation}
\label{v1v2}
V_1 [\psi ]\equiv
\left(\!\!
\begin{array}{c}
L_0\psi'(0^+) +i \psi(0^+)\\
-L_0\psi'(0^-)+i\psi(0^-)
\end{array} \!\!
\right)\; ; \quad
V_2 [\psi ]\equiv
\left(\!\!
\begin{array}{c}
-L_0\psi'(0^+) +i \psi(0^+)\\
L_0\psi'(0^-)+i\psi(0^-)
\end{array}\!\!
\right)
\end{equation}
the condition of probability current conservation, $j(0^+)=j(0^-)$, can be rewritten as the requirement that the vectors $V_1 [\psi ]$ and  $V_2 [\psi ]$ have the same length: $V_2^{\dagger }[\psi ] V_2[\psi ] = V_1^{\dagger } [\psi ]V_1 [\psi ]$. Thus, there must exist an unitary matrix $U$ such that \cite{FTs00}
\begin{equation} \label{vuv}
V_2 [\psi ]
= U V_1 [\psi ] \; .
\end{equation}
The boundary conditions (\ref{bc1})-(\ref{funct}) can be written in matrix form in terms of the vectors $V_1 [\psi ]$ and $V_2 [\psi ]$ as
\begin{equation}
\label{bcm}
\Omega =A_1V_1[\psi ]-A_2V_2[\psi ]\; , \\
\end{equation}
with
\begin{equation} \label{AA}
A_1=\frac{1}{2}
\left(
\begin{array}{ll}
-i&i\\
L_0^{-1}&L_0^{-1}
\end{array}
\right)\quad,\quad
A_2=\frac{1}{2}
\left(
\begin{array}{ll}
i&-i\\
L_0^{-1}&L_0^{-1}
\end{array}
\right),
\end{equation}
and after using the condition of probability current conservation, eq. (\ref{vuv}), the above expression for $\Omega$ becomes
\begin{eqnarray}
\label{solV1}
\Omega &=&\left[A_1-A_2\,U\right]\,V_1[\psi ]\\
\nonumber
\!\!\!\!\!\!\!\!\!\!\!\!\!\!\!\!&\!\!\!\!\!\!\!\!\!\!\!\!\!\!\!\!=\!\!\!\!\!\!\!\!\!\!\!\!\!\!\!\!&\frac{1}{2}
\left(
\begin{array}{ll}
-i\left[ 1+ e^{i\theta}\left(z+w^*\right)\right] & i\left[ 1+e^{i\theta}\left(z^*-w\right)\right] \\
L_0^{-1}\left[ 1-e^{i\theta}\left(z-w^*\right)\right] &
L_0^{-1}\left[ 1-e^{i\theta}\left(z^*+w\right)\right]
\end{array}
\right) V_1 [\psi ],
\end{eqnarray}
where we used the following parametrization for the unitary matrix $U$:
\begin{eqnarray}
\label{umatrix}
  U &\!\!=\!\!& e^{i \theta} \left(
                       \begin{array}{cc}
                         z & w \\
                         -w^* & z^* \\
                       \end{array}
                     \right), \quad \theta \in [0,\pi ), \quad |z|^2 + |w|^2 = 1.
\end{eqnarray}
After rewriting $V_1 [\psi ]$ in (\ref{solV1}) in terms of $\Phi(0^\pm)$ we obtain $\Omega $ in the form (\ref{funct}), with $M_{\pm}$ given by
\begin{eqnarray}
\label{solvPhi}
M_+&=&
\frac{1}{2}
\left(
\begin{array}{ll}
1+e^{i\theta}\left(z+w^*\right)& -iL_0\left[1+e^{i\theta}\left(z+w^*\right)\right]\\
iL_0^{-1}\left[1-e^{i\theta}\left(z-w^*\right)\right] &
\;\; 1-e^{i\theta}\left(z-w^*\right)
\end{array}
\right) \nonumber \\
\\
M_-&=&
\frac{-1}{2}
\left(
\begin{array}{ll}
-1-e^{i\theta}\left(z^*-w\right)& -iL_0\left[ 1+e^{i\theta}\left(z^*-w\right)\right]\\
iL_0^{-1}\left[1-e^{i\theta}\left(z^*+w\right)\right]&
-1+e^{i\theta}\left(z^*+w\right)
\end{array}
\right) \nonumber
\end{eqnarray}
Then, the b.c. (\ref{bcm}) can be rewritten as
\begin{equation} \label{tt}
R_+\Phi\left(0^+\right)=R_-\Phi\left(0^-\right),
\end{equation}
where we defined $R_{\pm}=1-M_{\pm}$. It is easy to show that $\det R_+=-w^*e^{i\theta}$ and that $\det R_-=we^{i\theta}$. Thus, $R_+$ and $R_-$ are \emph{both} invertible if, and only if, $w\neq 0$, in which case we can write
\begin{equation}
\label{lambda}
\Phi\left(0^+\right)=\Lambda \;\Phi\left(0^-\right),
\end{equation}
where
\begin{equation}
\label{deflambda}
  \Lambda =  R_+^{-1}R_-  = \frac{i}{w^*}  \left(
                            \begin{array}{cc}
                              \sin \theta - \Im(z) & L_0(\cos \theta + \Re(z)) \\
                             \frac{-\cos \theta + \Re(z)}{L_0} & \sin \theta + \Im(z) \\
                            \end{array}
                          \right)\, ,
\end{equation}
with $\Re(z)$ and $\Im(z)$ indicating the real and imaginary parts of $z$, respectively. From (\ref{deflambda}), it follows that $|\det \Lambda|= |-\frac{w}{w^*}|=1$ and $\Lambda$ can be rewritten in the form
\begin{equation}
\label{translink}
 \Lambda={\mathrm{e}}^{i\varphi}\left(
\begin{array}{cc}
a&b\\
c&d
\end{array}
\right), \quad ad-bc=1, \quad \varphi\in[0,\pi) \quad\mathrm{and} \quad a,b,c,d \in \mathbb{R}.
\end{equation}
The four parameter family of point interactions given by eq. (\ref{translink}) represents the most general point interactions consistent with a rigorous distributional approach to Schr\"{o}dinger equation, under the requirements ($i$)-($iii$) and the condition $\det R_{\pm} \neq 0$ [see (\ref{tt})]. This result is in agreement with the results obtained by the method of SAE, and corresponds to the case of a four parameter family of \emph{non-separated} solutions \cite{ADK98}.

From eqs. (\ref{bcm}) and (\ref{lambda}) in the case $w\neq 0$ one can rewrite the functionals $\alpha_1$ and $\alpha_0$ in terms of $\Phi\left(0^-\right)$ as
\begin{equation}
\label{functsols}
\Omega =\Phi(0^+)-\Phi(0^-) =\left(\Lambda-1\right)\Phi\left(0^-\right)\; ,
\end{equation}
from which one can obtain the explicit form of the interaction distribution as
\begin{eqnarray}
\nonumber
s\left[\psi\right](x)&=&\left[c\;{\mathrm{e}}^{i\varphi}\psi\left(0^-\right)+\left(d\;{\mathrm{e}}^{i\varphi}-1\right)\psi'\left(0^-\right)\right]\delta(x)\\
&+&\left[\left(a\;{\mathrm{e}}^{i\varphi}-1\right)\psi\left(0^-\right)+b\;{\mathrm{e}}^{i\varphi}\psi'\left(0^-\right)\right]\delta'(x) \, .
\label{intermexpl}
\end{eqnarray}
Alternatively, one can invert (\ref{lambda}) and substitute into (\ref{functsols}) to obtain
\begin{eqnarray}
s\left[\psi\right](x)&=&\left[c\;{\mathrm{e}}^{-i\varphi}\psi\left(0^+\right)-\left(a\;{\mathrm{e}}^{-i\varphi}
-1\right)\psi'\left(0^+\right)\right]\delta(x) \nonumber \\
&-&\left[ \left(d\;{\mathrm{e}}^{-i\varphi}-1\right)\psi\left(0^+\right)-b\;{\mathrm{e}}^{-i\varphi}\psi'\left(0^+\right)\right]\delta'(x) \, .
\label{interplus}
\end{eqnarray}

In order to complete the analysis of Schr\"odinger's equation with point interactions, we still have to consider the case in which $w=0$ (which reduces the number of free parameters to 2). In this case both $\det R_{\pm} = 0$ and, from eq. (\ref{vuv}), we obtain
\begin{eqnarray}
(z e^{i \theta}-1)\psi(0^+) -i L_0(z e^{i \theta}+1)\psi'(0^+) &=&0 ,
\label{w01} \\
(z^* e^{i \theta}-1)\psi(0^-) + i L_0(z^* e^{i \theta}+1)\psi'(0^-)&=&0. \label{w02}
\end{eqnarray}
The solution of these equations is as follows. If $z e^{i \theta} +1= 0$ ($z^* e^{i \theta} +1= 0$) we obtain $\psi(0^+)=0$ with $\psi'(0^+)$ an arbitrary finite number [$\psi(0^-)=0$ and $\psi'(0^-)$ arbitrary, respectively]. On the other hand, if $z
e^{i \theta} +1\neq 0$ it follows that $\psi'(0^+) = h_+ \psi (0^+)$, with $h_+= \frac{z e^{i \theta}-1}{i L_0(z e^{i
\theta}+1)}= \frac{2 \Im(z e^{i \theta})}{L_0 |1+ z e^{i \theta}|^2}$ a real number [analogously, if  $z^* e^{i \theta} +1\neq 0$, we have $\psi'(0^-) = h_- \psi (0^-)$ with $h_-= \frac{1-z^* e^{i \theta}}{i
L_0(z^* e^{i \theta}+1)}= \frac{-2 \Im(z^* e^{i \theta})}{L_0 |1+ z^* e^{i \theta}|^2}$ real]. Therefore, one can
summarize the case $w=0$ as
\begin{equation} \label{sep}
\psi'\left(0^{\pm}\right)=h^{\pm} \psi\left(0^{\pm}\right), \quad h^{\pm}\in \mathbb{R}\cup\{\infty\}\, ,
\end{equation}
which, for instance, implies that $j(0^+)=j(0^-)=0$. Again, this result is in agreement with the results obtained from the SAE approach, and corresponds to a two-parameter family of \emph{separated} solutions \cite{ADK98}. In this case the explicit form of the interaction distribution is
\begin{eqnarray}
\label{intimperm}
s[\psi](x)&=&\left[h_+\psi(0^+)-h_-\psi(0^-)\right]\delta(x) + \left[\psi(0^+)-\psi(0^-)\right]\delta'(x)\\
\nonumber
&=&\left[\psi'(0^+)-\psi'(0^-)\right]\delta(x) + \left[h_+^{-1}\psi'(0^+)-h_-^{-1}\psi'(0^-)\right]\delta'(x) \; .
\end{eqnarray}

The ability to provide explicit forms for the interaction, such as (\ref{intermexpl}), (\ref{interplus}) and (\ref{intimperm}), in terms of well defined distributions (consistent with the equation of motion), is the main difference between the results obtained from the distributional and the SAE approaches. While both methods coincide in the characterization of all point interactions by a four parameter family, only the distributional approach provides such an explicit visualization of the interaction as a distribution concentrated at the origin, a feature which allows one to investigate the properties of the interaction under symmetry transformations in a straightforward way. This will be used in the next section to properly define the notion of a point interaction that is odd under parity -- a non-trivial task in the context of SAE ( see, for instance, \cite{ADK98} for a broad investigation of the symmetries of the Schr\"odinger operators with point interactions).

%%%%%%%%%%%%%%%%%%%%%%%%%%%%%%%%%%%%%%%%%%%%%%%%%%%%%%%%%%%%%%%%%%%%%%%%%%%%%%%%%%%%%%%%%%%%%%%%%%%%%%%%%%%%%%%%%%%%%

\subsection{Symmetry under parity transformations}
\label{symm}

Let us investigate the behavior of the Schr\"odinger equation (\ref{Ss}) under parity transformations, in order to select subfamilies of interactions according to their symmetry properties.  From (\ref{sp}) and (\ref{solV1}) the interaction distribution can be put in the form
\begin{eqnarray}
\nonumber
s[\psi](x)&=&
\left(
\begin{array}{lr}
\delta'(x)\, ,&\delta(x)
\end{array}
\right)\,
\Omega \\
\label{intpsi}
&=&\left(
\begin{array}{lr}
\delta'(x) \, ,&\delta(x)
\end{array}
\right) \,\left[ A_1-A_2\;U\right] \,V_1[\psi],
\end{eqnarray}
with the matrices $A_i$'s given by (\ref{AA}). Defining $\phi(-x)\equiv \psi(x)$ and noticing that $\psi(0^\pm)=\phi(0^\mp)$ and $\psi'(0^\pm)=-\phi'(0^\mp)$, the Schr\"odinger equation (\ref{Ss}) can be rewritten in terms of $\phi$ as
\begin{equation}
\phi''(-x)+k^2\phi(-x)=\left(
\begin{array}{lr}
\delta'(x)\, ,&\delta(x)
\end{array}
\right)\,\left[ A_1-A_2\;U\right] \,\sigma_1\,V_1[\phi],
\end{equation}
where $\sigma_i$ ($i=1,2,3$) indicates a Pauli matrix and  $V_1[\phi]$ is obtained from (\ref{v1v2}) by making the substitution $\psi\to \phi$. Performing a space reflection, $x\to -x$, and taking into account that $\delta'(-x)=-\delta'(x)$ and $\delta(-x)=\delta(x)$ the above equation becomes
\begin{eqnarray}
\nonumber
\phi''(x)+k^2\phi(x)&=&\left(
\begin{array}{lr}
\delta'(x)\, ,&\delta(x)
\end{array}
\right)\,\left[ -\sigma_3\left(A_1-A_2\;U\right)\sigma_1\right] \,V_1[\phi]\\
\label{transpareq}
&\equiv& \tilde{s}[\phi](x).
\end{eqnarray}
Since $-\sigma_3\,A_j\sigma_1=A_j$, ($j=1,2$), and
\begin{equation}\label{Utilde}
\tilde{U}=\sigma_1U\sigma_1= e^{i\theta}
\left(
\begin{array}{cc}
z^*&-w^*\\
w&z
\end{array}
\right),
\end{equation}
we have that
\begin{equation}\label{condpar}
-\sigma_3\left(A_1-A_2\;U\right)\sigma_1 =A_1-A_2\,\tilde{U}
\end{equation}
and, therefore, $\phi$ solves a Schr\"odinger equation with an interaction term $\tilde{s}[\phi](x)$ which conserves probability across the origin and it is completely determined by the unitary matrix $\tilde{U}$.

For a regular potential, the interaction is said to be \emph{even} under parity if $\phi(x)=\psi(-x)$ is a solution of the \emph{same} Schr\"odinger equation solved by $\psi(x)$. Extending this concept to point interactions, we say that an interaction is \emph{even} if $\tilde{s}[\phi](x)=s[\phi](x)$, which is equivalent to the condition $\tilde{U}=U$. From (\ref{Utilde}) we see that this condition is satisfied if, and only if, $z=z^*$ and $w=-w^*$.  Thus, for $w\neq 0$ equations (\ref{deflambda}) and (\ref{translink}) imply that the interaction is even if  $a=d$ and $\varphi=0$, whereas for $w=0$ we must have $h_+=-h_- <\infty$ or $h_+=h_-=\infty$.

As an example of an even interaction we mention the well-known $\delta$-interaction, given by $\varphi =0, a=d=1$, $b=0$ and $c$ an arbitrary real number. Another example of an even interaction is given by the choice $\varphi=0$, $a=d=1$, $c=0$ and $b$ an arbitrary real number, which corresponds to the interaction term $s[\psi](x)=b\psi'(0^-)=b\psi'(0^+)\equiv b\psi'(0)$, the so-called $\delta'$-interaction -- which, as it is well-known, does not have the ``correct" (odd) symmetry under parity, as one would expect from an interaction associated to the ``potential" $\delta'$ (see, for instance, \cite{CNP97,RTa96, CSh98}).

Given the ``unexpected" behavior of the $\delta'$-interaction under parity transformations, let us investigate whether the four parameter family of general point interactions includes or not any interaction which behaves under parity in the same way that we would expect of an interaction associated with an odd potential. To this end, we recall that for a regular odd potential $V(x)$ [i.e., $V(-x)=-V(x)$] it follows that $\phi(x)=\psi(-x)$ satisfies a Schr\"odinger equation for an interaction term with the \emph{opposite} sign with respect to the interaction term in the equation satisfied by $\psi (x)$. This is generalized to point interactions by saying that an interaction is \emph{odd} under parity if it satisfies the condition $\tilde{s}[\phi](x)=-s[\phi](x)$. From (\ref{intpsi}), (\ref{transpareq}) and (\ref{condpar}) this condition is satisfied if, and only if,
\begin{equation}\label{odd}
A_1-A_2\,\tilde{U}=-\left[ A_1-A_2\,U\right] .
\end{equation}
Condition (\ref{odd}) implies that $\Re{(z)}=0, \theta=\frac{\pi}{2}$ and $\Im{(w)}=-1$;  these, together with the unitary condition (\ref{umatrix}), give $\Im{(z)}=\Re{(w)}=0$. Notice that this case ($w=-i\neq 0$) is characterized by a matrix $\Lambda$ which, according to (\ref{deflambda}), is the identity matrix. This corresponds to an interaction term which is identically zero and proves that \emph{there is no point interaction in non-relativistic one-dimensional quantum mechanics that behaves properly as an odd interaction}. Therefore, point interactions are either even or have no defined parity under space reflections.

A similar result about the nonexistence of a ``genuine" $\delta'$-interaction was obtained in \cite{RTa96} through a \emph{particular} regularization procedure. We stress that our result is mathematically rigorous and \emph{independent of regularization}, thus helping to clarify and to put into perspective any result obtained from a regularization procedure. In fact, suppose that a regularized product $\psi(x)V_\epsilon(x)$ is used as the interaction term into the Schr\"odinger equation, where $V_\epsilon$ is a sequence of regular distributions converging to a point potential $V(x)$ in the zero-range limit  $\epsilon \rightarrow 0$ (here $\epsilon$ may indicate multiple indices). If in this limit the regular products $\psi(x)V_\epsilon(x)$ converge in the sense of distributions, our approach implies that the limiting distribution must be a member (or a subfamily) of the four-parameter family of point interactions described above, which does not contain any genuinely odd interaction. Thus, we conclude that there is no regularization procedure which is able to produce an odd point interaction: even if all the regular products $\psi(x)V_\epsilon(x)$ of the sequence are odd under parity, in the zero-range limit the interaction term will lose this property. As an illustration of this fact, we recall that in \cite{Seb86} \v{S}eba demonstrated that if one considers the $\delta'(x)$ potential as the zero-range limit of the particular sequence of (singular) distributions $V_\epsilon(x)=\left[ \delta(x+\epsilon)-\delta(x-\epsilon)\right]/2\epsilon$ (for which the product $\psi(x)V_\epsilon(x)$ is well defined for all $\epsilon>0$), the interaction term obtained in the zero range limit corresponds to an impenetrable barrier, with reflection and transmission coefficients $R=-1$ and $T=0$ (see also \cite{CNP97})\footnote{\v{S}eba \cite{Seb86} has also shown that the same result holds if one uses as $V_\epsilon(x)$ a sequence of infinitely smooth functions converging to $\delta' (x)$ in the sense of distributions.}. Notice that for $\epsilon>0$ each regularized product $\psi(x)V_\epsilon(x)$ behaves as an \emph{odd} interaction, but in the limit $\epsilon\to 0$ these products converge to an interaction of the form (\ref{intimperm}) with $h_+=h_-=\infty$, which corresponds to an impenetrable $\delta$-interaction -- an \emph{even} interaction under parity transformations.
%

%%%%%%%%%%%%%%%%%%%%%%%%%%%%%%%%%%%%%%%%%%%%%%%%%%%%%%%%%%
\section{Dirac's equation with point interactions}
\label{R}

In this section we generalize the approach introduced in the previous section to find the most general family of point interactions allowed in one-dimensional relativistic quantum mechanics, in the context of Dirac's equation. Let us first recall that, in one dimension, the time-independent Dirac equation for a particle of mass $m$ interacting with a \emph{regular} potential $V(x)$ is (throughout this section we adopt natural units, $\hbar =c=1$)
\begin{equation} \label{dir}
\left( i \alpha_x \frac{d}{d x} - \beta m +E \right) \psi (x) = \psi (x) V(x),
\end{equation}
where $\alpha_x$ and $\beta$ are the Dirac matrices satisfying  $\{\alpha_x , \beta\} = 0$ and which can be chosen
as $\beta = \sigma_3$ and $\alpha_x = \sigma_1$, with $\sigma_i$ indicating the Pauli's matrices. The Dirac spinor $\psi (x)$ can be written in terms of its two components as $ \psi (x) = \left (u(x), \; v(x) \right)^T$, where $T$ stands for the transpose.

As it is well known, in the relativistic case even for the simplest point interaction (i.e., the Dirac delta potential) the interaction term cannot be naively defined as the product $\psi (x) V(x)$: defining the interaction as $\psi(0)\delta(x)$ is inconsistent with the fact that Dirac's equation implies that $\psi(x)$ must be discontinuous at the origin, as can be seen by equating the singular order in both sides of the resulting equation. Thus, in the same spirit of the previous section, we rewrite Dirac's equation (\ref{dir}) with the interaction term substituted by a well-defined interaction distribution,
\begin{equation} \label{dirD}
\left(i \alpha_x \frac{d}{d x} - \beta m  +E\right) \psi (x) = S[\psi](x) \; ,
\end{equation}
where $S[\psi](x)$ is a two-component spinor whose components are well defined distributions. Similarly, we demand that requirement $(i)$  and $(ii)$ of the previous section be satisfied. Together with Theorem 1  these imply that
\begin{equation} \label{S}
S[\psi](x)= \sum_{n=0}^{r_s} \, \Omega_n[\psi]\, \delta^{(n)}(x)\; ,
\end{equation}
where $r_s$ is the singular order of the interaction term and $\Omega_n[\psi]= (\varphi^{0}_n[\psi], \; \varphi^{1}_n[\psi] )^T$ are spinors whose (yet to be determined) components are linear functionals of the boundary values $\psi(0^\pm)$,  which completely characterize the interaction.  However, the most singular point interaction  allowed in the one dimensional Dirac's equation  (which still gives the possibility to build normalizable wave functions as superpositions of the stationary solutions) is described by a distribution with singular order zero, corresponding to a $\delta$ interaction. Thus, from now on we restrict ourselves to consider equation (\ref{S}) with $r_s=0$,  which is then reduced to
\begin{equation}\label{intermD}
S[\psi](x)=
\Omega_0[\psi ]\,\delta(x)\; ;\quad \quad \Omega_0 [\psi ] =\left( \!
\begin{array}{c}
\varphi^{0}\\
\varphi^{1}
\end{array} \!
\right)\, ,
\end{equation}
where we dropped the subscript ``$0$" in the components $\varphi^{0(1)}$, to simplify the notation.

Substituting the interaction (\ref{intermD}) into Dirac's equation results in b.c. for the Dirac spinor given by
\begin{equation} \label{bcspinor}
\psi(0^+)-\psi(0^-)=-i\alpha_x   \Omega_0 [\psi ]
\end{equation}
as can be seen by taking an indefinite integral of the Dirac equation and using the fact that the primitive $\psi^{(-1)}$ is continuous (since $r_s=0$, this primitive has singular order $-2$).

The equation resulting from the substitution of (\ref{intermD}) into (\ref{dirD}) can be rewritten as two coupled first order equations for $u(x)$ and $v(x)$, which in turn can be reduced to
\begin{equation} \label{v}
v(x) = \frac{1}{(E+m)} \left[ \varphi^1 \delta (x) - i u'(x) \right]\; ,
\end{equation}
and a second order equation for $u(x)$:
\begin{equation}
\label{u}
u''(x) + k_r^2u(x)=(E+m)\varphi^{0}\,\delta(x)-i \varphi^{1}\,\delta '(x)\;,
\end{equation}
with $k_r\equiv \sqrt{E^2-m^2}$. Equation (\ref{u}) is a Schr\"{o}dinger-like equation for $u(x)$ with a singular interaction of order $+1$.

The solution of (\ref{v})-(\ref{u}) follows directly from the results in section \ref{NR}. However, before proceeding to obtain the explicit form of the interaction term, let us first consider the requirement $(iii)$ of probability current conservation. In one dimension the relativistic current density is given by $j^{1} (x) = \psi^{\dagger}(x)\alpha_x\psi (x) = u^*(x)v(x)+u(x)v^*(x)$; from eq. (\ref{v}) it follows that $v(0^{\pm}) = - \frac{i}{(E+m)}u'(0^{\pm})$ and therefore
\begin{equation}
j^{1} (0^{\pm}) = \frac{1}{i(E+m)}\left\{ u^* (0^{\pm})u'(0^{\pm})- {u^*}'(0^{\pm})u(0^{\pm}) \right\}\, .
\end{equation}
Hence, conservation of the ``non-relativistic" probability current associated with $u(x)$ implies the conservation of the relativistic current across the singular point.

The formal similarity between (\ref{u}) and the Schr\"odinger equation with the interaction term (\ref{sp}) allows us to find the b.c. for $u$ and $u'$ at the origin by identifying $(E+m)\varphi^{0}=\alpha_0$ and $i \varphi^{1}=\alpha_1$, so that all the results from the previous section are also valid for the $u$-component of the Dirac spinor. Then, by taking into account (\ref{v}) we obtain the b.c. for $v$ and for the Dirac spinor $\psi$. In fact, it follows from (\ref{tt}) that the vector $\Upsilon (x) \equiv \left( u(x), \;u'(x)\right)^T $ satisfies the b.c. $R_+ \Upsilon(0^+) = R_- \Upsilon (0^-)$, with $R_{\pm}$ given in section \ref{NR}. Noticing that  $\Upsilon(0^{\pm}) = G \psi (0^{\pm})$, with the $G$ matrix defined as
$$
G=\left(
            \begin{array}{cc}
              1 & 0 \\
              0 & i(E+m) \\
            \end{array}
          \right)\, ,
$$
from the results for the non-relativistic theory it follows directly that:
\begin{itemize}
    \item[(\textit{a})] If $\det R_{\pm} \neq 0$ the relativistic b.c. can be written as
        \begin{equation}
        \label{lambdefrel}
        \psi (0^+) = \Lambda_r \,\psi (0^-)\, ,
        \end{equation}
        where $\Lambda_r = G^{-1} \Lambda G$ can be written as (the subscript ``$r$" is to indicate relativistic quantities)
        \begin{equation}
         \label{lbdRel}
          \Lambda_r ={\mathrm{e}}^{i\varphi_r}\left(
         \begin{array}{cc}
         a_r&ib_r\\
         -ic_r&d_r
         \end{array} \right), \quad a_rd_r-b_rc_r=1,
        \end{equation}
        with $\varphi_r\in[0,\pi)$ and $a_r,b_r,c_r,d_r \in \mathbb{R}$ are (dimensionless) constants.

    \item[(\textit{b})] If $\det R_{\pm} = 0$, the relativistic b.c. can be written as
        \begin{equation} \label{sepR}
         v \left(0^{\pm}\right)=i h_r^{\pm} u\left(0^{\pm}\right), \quad h_r^{\pm}\in \mathbb{R}\cup\{\infty\}\, .
        \end{equation}
\end{itemize}

Therefore, we conclude that, similarly to the non-relativistic case, there are two possible situations in one-dimensional relativistic quantum mechanics: for \emph{non-separated} solutions [case $(a)$ above] there is a four-parameter family of interactions that includes all the possible point interactions, and for \emph{separated} solutions [case $(b)$ above; corresponding to set $w=0$ in (\ref{umatrix})] the number of free parameters is reduced to two. Again, this result agrees with that obtained by SAE \cite{BDa94} -- for works dealing with particular one-parameter subfamilies of these interactions see \cite{SMa81, MSt87, DAM89, Roy93}.

The explicit form of the relativistic interaction term for the case ($a$) above can be found from (\ref{intermD}), (\ref{bcspinor}) and (\ref{lambdefrel}) as
\begin{equation} \label{intR}
S[\psi](x) = i\alpha_x \left( \Lambda_r -1 \right)\psi(0^-)\,\delta(x),
\end{equation}
whereas for the case ($b$) we use (\ref{intermD}), (\ref{bcspinor}) and (\ref{sepR}) to obtain
\begin{equation}\label{explsepR}
S[\psi](x) = i\alpha_x \left[
\left(
\begin{array}{ll}
1&0\\
i h_r^+&0
\end{array}
\right)\psi(0^+) -
\left(
\begin{array}{ll}
1&0\\
i h_r^-&0
\end{array}
\right)\psi(0^-)
\right]\,\delta(x).
\end{equation}

It is convenient to express (\ref{u}) explicitly in terms of the parameters in (\ref{lbdRel}) or (\ref{sepR}). For case $(a)$, with the interaction term given by (\ref{intR}), the $u$-component of the Dirac spinor satisfies [also see (\ref{intermD})]
\begin{eqnarray}
\nonumber
u''(x)+k_r\,u (x)&=&\left[(E+m)c_r\,\mathrm{e}^{i\varphi_r}u(0^-) + \left(d_r\mathrm{e}^{i\varphi_r}-1 \right)u'(0^-) \right]\delta(x)\\
\label{ueqR}
&+& \left[ \left(a_r\mathrm{e}^{i\varphi_r}-1 \right)u(0^-)   + \frac{b_r}{E+m}\mathrm{e}^{i\varphi_r}u'(0^-)  \right]\delta'(x)\, .
\end{eqnarray}
Similarly, when the interaction term is given by (\ref{explsepR}) [case (b)] the equation for the $u$-component reads
\begin{eqnarray}
\nonumber
u''(x)+ k^2_r u (x) &=& (E+m)\left[-h_r^+\,u(0^+) + h_r^-\,u(0^-)\right]\delta(x)\\
\label{ueqRsep}
&& + \left[u(0^+)-u(0^-)\right]\delta'(x).
\end{eqnarray}
Equations (\ref{ueqR}) and (\ref{ueqRsep}) are suitable to analyze the non-relativistic limit (characterized, as usual, by $E\to m$, $k_r\to k$) of Dirac's equation. Following a procedure analogous to the one presented in \cite{BDr64}, it can be shown that in the non-relativistic limit $v(x)$ is the ``small" component of the Dirac spinor, whereas the above Schr\"odinger-like equations satisfied by the ``large" component $u(x)$ correspond, in that limit, to equations (\ref{intermexpl}) and (\ref{sep}), respectively. Comparing the above equations with their equivalent in the non-relativistic case [see eqs. (\ref{intermexpl}) and (\ref{intimperm}), respectively], we can establish a one-to-one correspondence between the parameters of the relativistic and non-relativistic families of point interactions:
\begin{eqnarray}
\label{rnrR}
&&\varphi=\varphi_r,\quad a=a_r,\quad b=\frac{b_r}{2m},\quad c=2m\,c_r,\quad d=d_r,\\
\label{rnrRsep}
&&h_\pm=-2m\,h_r^\pm\, ,
\end{eqnarray}
which was also reported in the context of SAE \cite{BDa94}. For instance, from these relations it is straightforward to see that the relativistic interaction that corresponds in the non-relativistic limit to a $\delta$-interaction with strength $\gamma$ is given by $\varphi_r=b_r=0$, $a_r=d_r=1$, $c_r=\gamma$, which corresponds to the following interaction term in the Dirac equation
\begin{equation}
S[\psi](x) = \frac{1}{2}\gamma ({\mathbf{1}}+\beta)\psi (0^-) \delta (x)\, ,
\end{equation}
where ${\bf 1}$ is the $2\times 2$ identity matrix. This interaction is an equal mix of electrostatic and scalar potentials, which is known to be a confining interaction \cite{DAM89} (i.e., the transmission amplitude approaches zero in the limit $\gamma \to \infty$). On the other hand, the relativistic interaction given by $\varphi_r =0$, $a_r=d_r = 1$, $b_r=-\gamma$ and $c_r=0$,
\begin{equation}
S[\psi](x) = \frac{1}{2}\gamma ({\bf 1}-\beta)\psi (0^-) \delta (x),
\end{equation}
is the inverted mix of electrostatic and scalar potentials (also confining \cite{DAM89}) and corresponds, in the non-relativistic limit, to the interaction given by $\varphi=c=0$, $a=d=1$ and $b=-\frac{\gamma}{2m}$, which is the so-called $\delta' $-interaction (with strength $-\gamma/4m^2$). Other well-known one-parameter subfamilies of relativistic interactions are the pure scalar ($\varphi_r=0$, $b_r=c_r=\gamma$, $a_r=d_r=1$) and pure electrostatic ($\varphi_r=0$, $-b_r=c_r=\gamma$, $a_r=d_r=1$), which do not correspond in the non-relativistic limit either to a $\delta$ or to the $\delta '$ interaction, as already observed by Benvegn\`{u} and D\c{a}browski \cite{BDa94}. These are, of course, just some examples of particular one-parameter subfamilies often studied in the framework of relativistic quantum mechanics in one dimension; all of them are included in the four-parameter family of interactions considered above.

Finally, the results of subsection \ref{symm}, concerning the characterization of non-relativistic point interactions with respect to their behavior under space reversal (parity transformations), can be straightforwardly extended to relativistic interactions. Accordingly, if $\psi(x)$ solves a Dirac equation with the interaction $S[\psi](x)$, the interaction is said to be \emph{even} under space reversal if $\tilde{\psi}(x)\equiv\beta \psi(-x)$ solves the same equation, and it is said to be \emph{odd} if $\tilde{\psi}(x)$ solves the Dirac equation for an interaction term with \emph{reversed} sign, $-S[\tilde{\psi}](x)$. The conditions that the relativistic parameters must satisfy to characterize an even interaction are $\varphi_r=0$, $a_r=d_r$ for the non separated solutions, and $h_r^+=-h_r^-<\infty$ or $h_r^+=h_r^-=\infty$ for the separated solutions. These are exactly the conditions that we would obtain from the correspondence relationships between the non-relativistic and relativistic parameters, eqs. (\ref{rnrR}) and (\ref{rnrRsep}). In a similar way, we conclude that there is no point interaction which is odd under space reversal in one dimensional relativistic quantum mechanics.

%%%%%%%%%%%%%%%%%%%%%%%%%%

\section{Concluding Remarks}
\label{concl}

In this work we considered a distributional approach to treat singular point potentials in one-dimensional quantum mechanics. The method relies on the fact that any singular distribution concentrated at a point must be a linear combination of the $\delta$-distribution and its derivatives (see Theorem 1). This, together with requirement $(ii)$ (section \ref{NR}), allows us to substitute the ill-defined product $V(x)\psi(x)$ in the Schr\"odinger and Dirac equations by a well-defined interaction distribution with a given singular order. Then, the interaction distribution is completely determined by imposing the physical requirement that the probability current is conserved across the singularity -- notice that even though the probability current is, in general, ill-defined \emph{at} the singularity, it is well-defined on both sides of the singularity, so that in all the steps we only deal with well defined distributions. Implicit in the method is a \emph{locality condition}, i.e., the interaction term depends both on singular potential and the wave function at the vicinities of the singular point.

The application of the distributional approach to both non-relativistic and relativistic quantum mechanics leads to a four-parameter family of point interactions, a result in agreement with that obtained by the method of self-adjoint extensions. It is important to notice that, while our results coincide with those of SAE in what regards the characterization of the family of all point interactions for Schr\"{o}dinger's and Dirac's theories, there is an important conceptual difference between the two methods. While in SAE the interaction is defined only in terms of the boundary conditions around the singular point, in the present distributional approach  the interaction is \emph{explicitly constructed} from simple mathematical and physical requirements as an \emph{interaction distribution} concentrated at the origin, which implies the boundary conditions satisfied by the wave function.

The fact that the interaction term is explicitly provided [following from the requirements $(i)$-$(iii)$ in section \ref{NR}] makes it particularly simple to analyze the symmetries of the interaction, as the investigation of parity transformations in sections \ref{NR} and \ref{R} shows. As one of the main results of the distributional approach such an analysis shows, in a simple way, why there is no \emph{odd} point interaction in one dimension. Indeed, an immediate general consequence of the distributional method is that any regularization using $\delta '$-converging sequences of functions for the potential must converge, \emph{in the sense of distributions}, to an interaction distribution \emph{which cannot be odd} under parity. We conjecture that a failure to satisfy this criterion is at the origin of the sometimes contradictory results appearing in the literature of regularization based treatments of zero-range interactions. Notice that, in general, these regularization methods assume from the start odd sequences of short-range potentials with the expectation of obtaining an ``odd $\delta '$-interaction" in the zero-range limit (even though, of course, this is no guarantee of convergence to an odd distribution, as clearly demonstrated by \v{S}eba \cite{Seb86}).

Finally, we note that the distributional approach to point interactions developed here offers a mathematically rigorous and less abstract  alternative to the (also rigorous) SAE method. In addition, the method is expected to generalize in a straightforward way to higher dimensions as well as to more general interactions in one or more dimensions which still present open questions, such as the Coulomb potential \cite{OVe09}. Preliminary work applying the distributional approach to the one-dimensional Coulomb interaction \cite{CLM04} suggests that the fact that one obtains an explicit form for the interaction distribution, together with symmetry considerations, may help to clarify some ambiguities concerning a proper characterization of this interaction.

%%%%%%%%%%%%%%%%%%%%%%%%%%%%%%%%%%%%%%%%%%%
%%%%%%%%%%%%%%%%%%%%%%%%%%%%%%%%%%%%%%%%%%%
\section*{Acknowledgments}

WM thanks Funda\c{c}\~ao Arauc\'aria for financial support. LAM thanks the Office of Undergraduate
Research, Concordia College, for partial financial support. JTL thanks F. M. de Andrade for stimulating discussions at the beginning of this work.

%%%%%%%%%%%%%%%%%%%%%%%%%%%%%%%%%%%%%%%%%%%%%

%\appendix
%\setcounter{section}{1}

%%%%%%%%%%%%%%%%%%%%%%%%%%%%%%%%%%%%%%%%%%%%%%

\end{document}